\pdfoutput=1
\documentclass[10pt, conference]{IEEEtran}
 \IEEEoverridecommandlockouts

\usepackage{cite}
\ifCLASSINFOpdf
\else
\fi
\usepackage{color}
\usepackage{graphicx}
\usepackage{url}
\usepackage{subfig}
\usepackage{fixltx2e}
\usepackage{listings}
\usepackage{fixltx2e}
\definecolor{lightgray}{rgb}{.9,.9,.9}
\definecolor{darkgray}{rgb}{.4,.4,.4}
\definecolor{purple}{rgb}{0.65, 0.12, 0.82}

\lstdefinelanguage{JavaScript}{
  keywords={typeof, new, true, false, catch, function, return, null, catch, switch, var, if, in, while, do, else, case, break},
  keywordstyle=\color{blue}\bfseries,
  ndkeywords={class, export, boolean, throw, implements, import, this},
  ndkeywordstyle=\color{darkgray}\bfseries,
  identifierstyle=\color{black},
  sensitive=false,
  comment=[l]{//},
  morecomment=[s]{/*}{*/},
  commentstyle=\color{purple}\ttfamily,
  stringstyle=\color{red}\ttfamily,
  morestring=[b]',
  morestring=[b]"
}

\newcommand{\code}[1]{\texttt{\small#1}}
\newcommand{\cell}[1]{\subsection*{#1}}
\newcommand{\pie}[1]{\smallskip\noindent\textbf{#1}}

\begin{document}
\title{Statically Checking Web API Requests in JavaScript}

\author{\IEEEauthorblockN{Erik Wittern*, Annie T. T. Ying*, Yunhui Zheng*, Julian Dolby, Jim A. Laredo}
\IEEEauthorblockA{IBM T. J. Watson Research Center, Yorktown Heights, NY, USA\\
Email: \{witternj, aying, zhengyu, dolby, laredo\}@us.ibm.com}
\thanks{* The author names were sorted alphabetically. The authors contributed equally to the work.}
}


%Telephone: (800) 555--1212, Fax: (888) 555--1212}


\maketitle

% !TEX root = ./paper.tex

\begin{abstract}

Many JavaScript applications perform HTTP requests to web APIs, relying on the request URL, HTTP method, and request data to be constructed correctly by string operations.
Traditional compile-time error checking, such as calling a non-existent method in Java, are not available for checking whether such requests comply with the requirements of a web API.
In this paper, we propose an approach to statically check web API requests in JavaScript.
Our approach first extracts a request's URL string, HTTP method, and the corresponding request data using an inter-procedural string analysis, and then checks whether the request conforms to given web API specifications.
We evaluated our approach by checking whether web API requests in JavaScript files mined from GitHub are consistent or inconsistent with publicly available API specifications. 
From the $6575$ requests in scope, our approach determined whether the request's URL and HTTP method was consistent or inconsistent with web API specifications with a precision of $96.0\%$.
Our approach also correctly determined whether extracted request data was consistent or inconsistent with the data requirements with a precision of $87.9\%$ for payload data and $99.9\%$ for query data.
In a systematic analysis of the inconsistent cases, we found that many of them were due to errors in the client code.
The here proposed checker can be integrated with code editors or with continuous integration tools to warn programmers about code containing potentially erroneous requests.

\end{abstract}


% \fi
\IEEEpeerreviewmaketitle

% !TEX root = ./paper.tex

\section{Introduction}
\label{sec:intro}
\begin{figure}%[htb]
  \centering
  \includegraphics[width=\columnwidth]{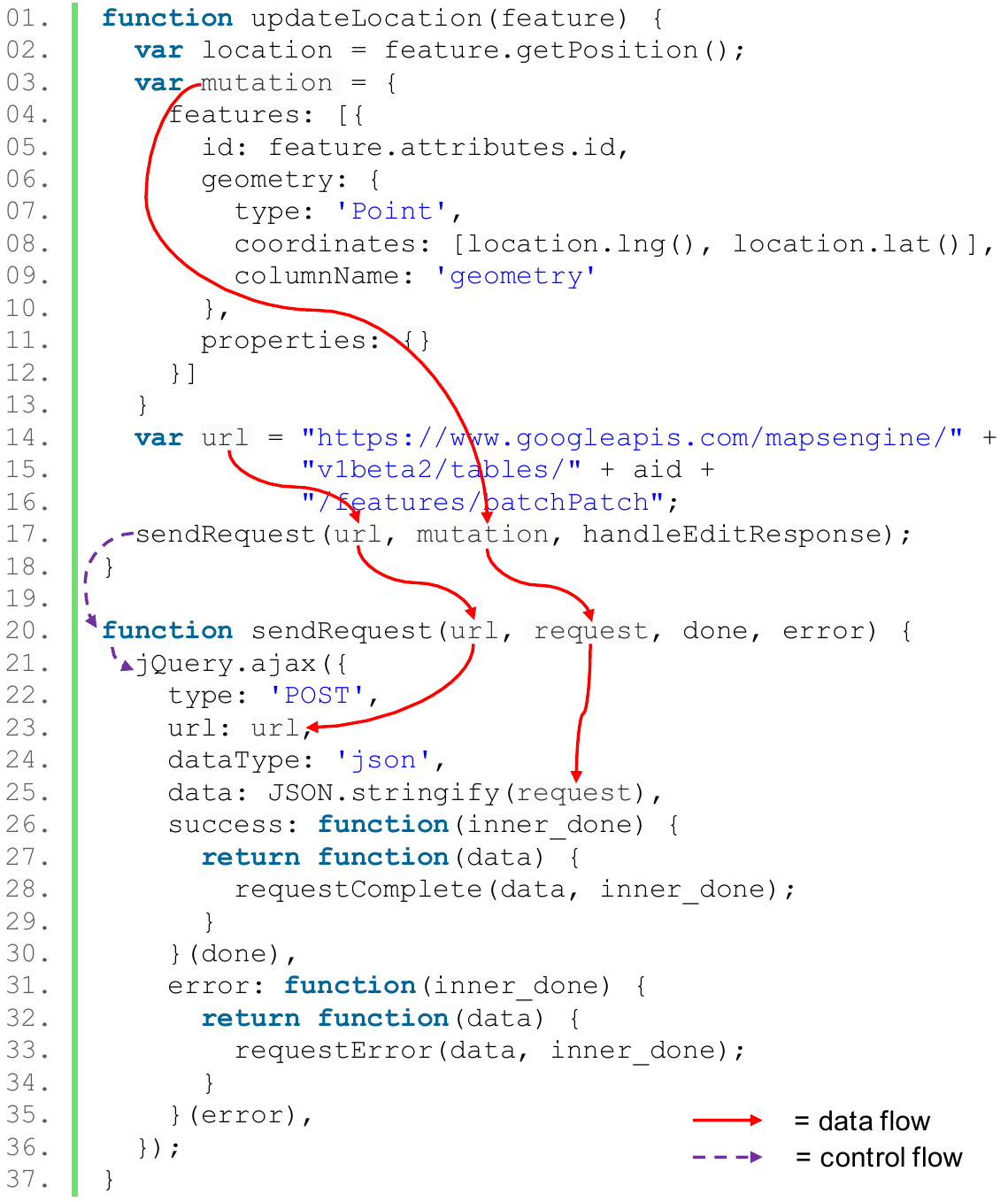}
  \caption{Code excerpt of a request to the Google Maps Engine API}
  \label{fig:example_inter_procedual}
\end{figure}

% Context: The role of web APIs:
Programmers write applications using a growing variety of publicly accessible web services. Catalogs such as IBM's API Harmony~\cite{APIHarmony,Wittern:2016}, Mashape's PublicAPIs~\cite{Mashape}, or ProgrammableWeb~\cite{ProgrammableWeb} list thousands of \emph{web Application Programming Interfaces} (web APIs) exposed by these services. Applications invoke web APIs by sending HTTP requests to a dedicated URL using one of its supported HTTP methods; required data is sent as query or path parameters, or within the HTTP request body. The URL, HTTP method, and data to send are all basically strings, constructed by string operations within the applications. Figure~\ref{fig:example_inter_procedual} shows an exemplary excerpt of such a JavaScript application performing these actions.
%In the example, the URL and data to be used in the request are defined in one function and then used to perform the request in another function. The red arrows indicate the flow of data, while the purple arrows indicate involved function calls.

% Problem: Errors appear only at runtime
When a request targets a URL that does not exist or sends data that does not comply with the requirements of the web API, a runtime error occurs. This prevalent calling mechanism for web APIs---which relies on a string URL, a HTTP method, as well as string input and output---does not allow type-safety checking. In other words, checks for traditional compile-time errors are not available for programmers writing code calling web APIs. A recent study found that a significant number of analyzed mobile applications will fail in light of changes to the web APIs they consume~\cite{Espinha:2015}. The situation is worsened as (web) applications are increasingly developed using dynamic languages like JavaScript, which generally also have minimal static checking.

As an example of resulting errors, we found code in GitHub that mistakenly attempts to make a request to \code{https://api.
spotify.com/v1/seach}, as opposed to invoking the correct URL ending with \code{/search}. 
Another example we found (Figure~\ref{fig:example_inter_procedual}) attempts to invoke the deprecated Google Maps Engine API.\footnote{\url{https://mapsengine.google.com/about/index.html}} 
A programmer wishing to avoid these errors can manually assess the correctness of web API requests by consulting the API's (online) documentation or formal web API specifications. Such specifications, like the OpenAPI Specification~\cite{OAI} (formerly known as \emph{Swagger}, the name we will use for the rest of the paper) can be created by API providers or third parties to document valid URLs, HTTP methods, as well as inputs and outputs that a web API expects.

% Solution approach:
Tools that can automate this manual---and thus error-prone and tedious---checking should have two desirable features:

\begin{enumerate}
  \item Such tools should statically analyze JavaScript source code to automatically identify HTTP requests and retrieve the related URL, HTTP method, and data, which are all encoded as strings and created using typical string operations like concatenation. In addition, the analysis must be inter-procedural as the strings can be assembled across functions.
  \item As input, such tools should make use of available specifications, like Swagger, for the definitions of valid URLs, HTTP methods, and data.  
\end{enumerate}
%Tools capable of detecting possible erroneous web API usage 
Such tools can report errors either real-time as a programmer is writing the application, or during continuous integration. In addition, they can help API providers to monitor usages of their APIs in publicly available code.

In this paper, with these two features in mind, we propose an approach that takes as input Swagger specifications and statically checks whether the web API requests in JavaScript code conform to these specifications. Our approach first extracts the URL string, HTTP method, and the corresponding data from a request, using an inter-procedural static program analysis capable of extracting strings~\cite{Feldthaus_icse2013}, and then checks whether the request conforms to publicly available web API specifications. For the initial implementation, we chose to handle requests written using the jQuery framework due to its popularity -- reportedly, ~70\% of websites use the jQuery framework~\cite{jQueryUsage}. The main contribution of our approach is in leveraging existing work in making static whole-program analysis possible for framework-based JavaScript web applications (i.e., ~\cite{DBLP:conf/pldi/SchaferSDT13,DBLP:conf/oopsla/AndreasenM14,DBLP:conf/kbse/KoLDR15,Feldthaus_icse2013}) and applying it to a new problem of checking whether a request is consistent with a web API specification.

We evaluated our approach by checking whether web API requests from over $6000$ JavaScript files on GitHub\footnote{\url{https://github.com/}} were consistent or inconsistent with publicly available web API specifications provided by the APIs Guru project~\cite{APIsGuru}. From $6575$ requests for which we had web API specifications available, our analysis achieved a precision of $96.0\%$ in correctly extracting and determining that the URL and HTTP method in a request is consistent or inconsistent with the corresponding web API specification. 
Our approach also correctly extracted and determined whether the request data was consistent or inconsistent with the data requirements with a precision of $87.9\%$ for payload data and $99.9\%$ for query data.
We systematically examined all the URLs and payload data that were inconsistent with any specification ($1477$ cases) and found that many of these inconsistencies were due to errors in the client code, including calls to deprecated APIs, errors in the URLs, and errors in data payload definitions. In only five of the $1477$ cases, limitations in our static analysis affected the matching of a request to a URL endpoint in the specification to the point of incorrectly flagging requests as mismatches. 
These limitations also extended to four out of $18$ cases where request payloads were mistakenly flagged as mismatches and two out of $41$ cases where query parameters were mistakenly flagged as mismatches. These results show that the static analysis is precise enough to be used in our proposed checker, for checking an application under development, or for checking the validity of web API usage in existing source code in case a web API undergoes changes.\footnote{Consider the high number of changes reported for various APIs at \url{https://www.apichangelog.com/}}

The remaining of this paper is organized as follows:
After illustrating an example (Section~\ref{sec:example}), we describe our approach: the static analysis (Section~\ref{sec:approach}) and the checker (Section~\ref{sec:checker}). We then present the evaluation (Section~\ref{sec:experiment}), related work (Section~\ref{sec:related-work}), threats (Section~\ref{sec:threats}), and conclusion (Section~\ref{sec:conclusion}).

% !TEX root = ./paper.tex

\section{Background and Example}
\label{sec:example}
In this section, we first introduce concepts and terms regarding web APIs and their specifications. We then demonstrate through an example the two steps of our approach: 
how we use static analysis to extract the strings constructing a web API request (Section~\ref{sub:example_static_analysis}) and how we check the results of the static analysis against Swagger specifications (Section~\ref{sub:example_checker}).

%\subsection{web APIs and web API Specifications}
%\label{sec:example_web_apis}
% Web APIs:
Web APIs are programmatic interfaces that applications invoke via HTTP to interact with remote \emph{resources}, such as data or functionalities. Resources are identified by URLs while the type of interaction (e.g., retrieval, update, deletion of a resource) depends on the HTTP \emph{method}. Following previous work, we refer to the combination of a URL and HTTP method as an API \emph{endpoint}~\cite{Suter:2015}. To be successfully invoked, some endpoints depend on additional data, for example the ID of a resource being sent as a path parameter within the URL or a new/updated state of a resource being sent in the body of an HTTP request.

\begin{figure}
\includegraphics[width=\columnwidth]{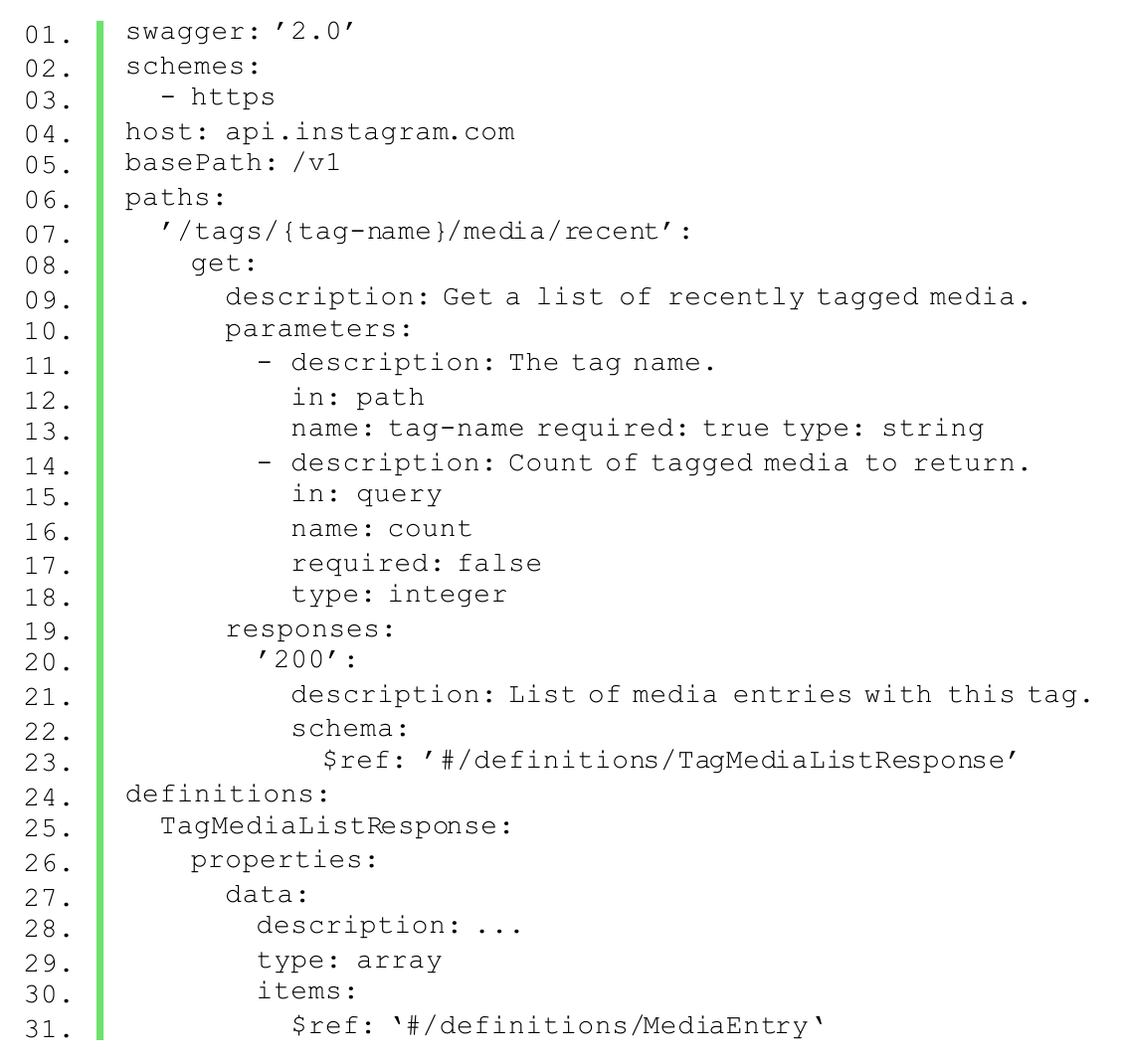}
\vspace{-5pt}
\caption{
\label{fig:instagram-swagger}Excerpt of a Swagger specification for the Instagram API, highlighting the \code{GET /tags/\{tag-name\}/media/recent} endpoint}
\end{figure}

% Formal web API specifications:
Application developers can learn the usage of the endpoints of an API either by consulting its online documentation, typically presented in HTML, or by relying on a formal web API specification. Specifications define, among other things, an API's endpoints as well as the data required for and returned by requests. The OpenAPI specification (Swagger) is one of these specifications, which enjoys broad industry support~\cite{OAI}. Figure~\ref{fig:instagram-swagger} shows an excerpt of a Swagger document describing the Instagram API. It defines, for example, the \code{schemes} of the API, its \code{host} and \code{basePath}, which together form the API's \emph{base URL}, in this case \code{https://api.instagram.com/v1}. Swagger defines the different endpoints of an API in the \code{paths} property, using URL templates (possibly containing path parameters, i.e., \code{\{tag-name\}} in the path \code{/tags/\{tag-name\}/media/recent}) and supported HTTP methods. Per endpoint, Swagger provides a human-readable \code{description}, definitions of the \code{parameters} (path and query parameters as well as required HTTP bodies), definitions of possible \code{responses}, as well as \code{security} requirements. Entries in the \code{definitions} property describe the structure of data to send to or receive from endpoints using JSONschema notation~\cite{JSONSchema} or a XML Object notation that is specific to Swagger. Data definitions can be referenced from endpoint definitions, as is exemplary shown for the \code{TagMediaListResponse} definition in Figure~\ref{fig:instagram-swagger}.

\subsection{Determining the content of a request in JavaScript code}
\label{sub:example_static_analysis}
The first step of the checker is to extract the specifics of web API requests in the code. Our approach employs an inter-procedural static analysis to extract URL and input strings from a web API request in JavaScript code. Recall the code in Fig.~\ref{fig:example_inter_procedual} as an example. Focusing on the \code{url} variable, we can see that it is composed from two constant strings and the \code{aid} variable in the function \code{updateLocation}. The value of \code{url} is then passed to \code{sendRequest}, where it flows into the \code{jQuery.ajax} call. The value of \code{aid} is a parameter and could be different in multiple runs. Hence, when we aim to extract the URL used in this request, we denote \code{aid} as a symbolic value \code{\{aid\}} using curly braces, indicating that the value is not known statically. Overall, the URL extracted for the shown request is \code{https://www.googleapis.com/mapsengine/ v1beta2/tables/\{aid\}/features/batchInsert}.

As this example shows, a simple textual search like \code{grep} would not be effective because the call site of the request~(e.g., \code{sendRequest} in Figure~\ref{fig:example_inter_procedual}) can be different from the definition of the URL string (\code{updateLocation} in Figure~\ref{fig:example_inter_procedual}). In addition, given a URL string can be assembled across multiple functions and lexical scopes (e.g., the URL \code{https://api.instagram.com/v1/tags/\{tag-name\}/me dia/recent} which our static analysis correctly extracts from the code excerpts in Figure~\ref{fig:analysis_moti}), resolving such an URL string requires non-trivial data flow analysis. The same holds for the HTTP method or request data values, which may be created within multiple functions.

\subsection{Checking a request against a web API Specification}
\label{sub:example_checker}
The second step of our approach checks whether the extracted information from the web API requests conforms with web API specifications.
Consider, for example, the URL \code{https://api.instagram.com/v1/tags/\{searchHashtag\}
/media/recent?client\_id=1e3...} extracted by our static analysis from the code excerpt in Figure~\ref{fig:analysis_moti}. Our check would start by determining whether it - together with the associated method - targets an actual endpoint defined in the Swagger specification of the Instagram API, including the \code{searchHashtag} path parameter. In addition, we can check whether the \code{client\_id} parameter is expected by the endpoint or if there are other query parameters required, which are missing in the URL. Finally, we can check if the data sent in the request body adheres to the data definitions in the Swagger specification.

Another option to check whether a web API request is correct would be to perform a dynamic analysis~\cite{Sen:2013:JSR:2491411.2491447}. However, invocations of web APIs often require authentication (for example, using API keys), so that a system using dynamic analysis would need to provision keys to register and even agree to the terms of service. Typically, terms of service are not easy to understand by a layman, and are much less likely to be encoded in a machine readable form to allow a program to decide whether or not to comply with the terms. Finally, even if the key provisioning issue was addressed, ensuring dynamic analysis has the proper code coverage is challenging.

%\input{background.tex}
% !TEX root = ./paper.tex

\section{Web API Usage Extraction}
\label{sec:approach}

\begin{figure}[t]
\centering
\includegraphics[width=\columnwidth]{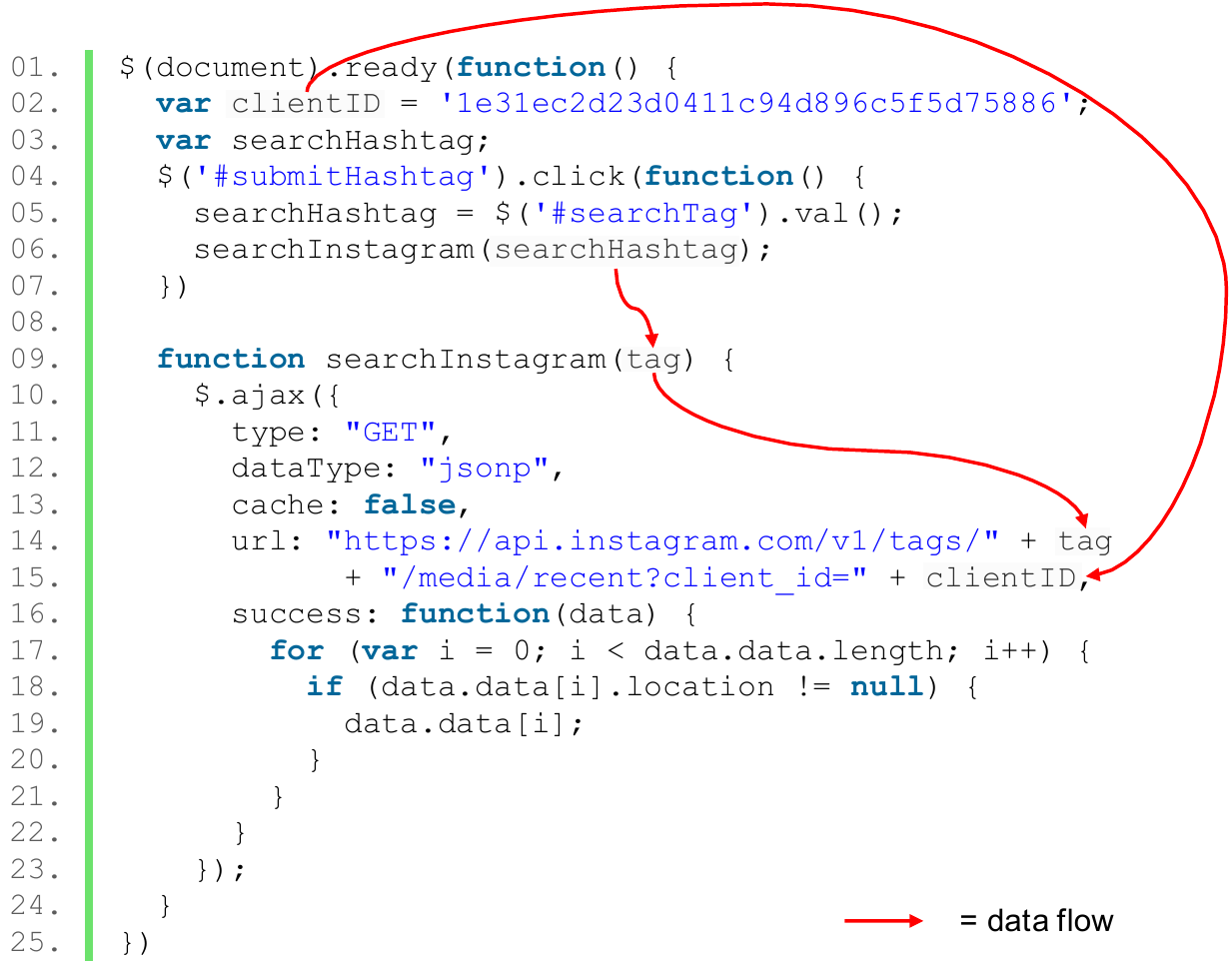}
\caption{
  Code excerpt of a request to the Instagram API}
\label{fig:analysis_moti}
\end{figure}

\begin{figure}[t]
\centering
\includegraphics[width=\columnwidth]{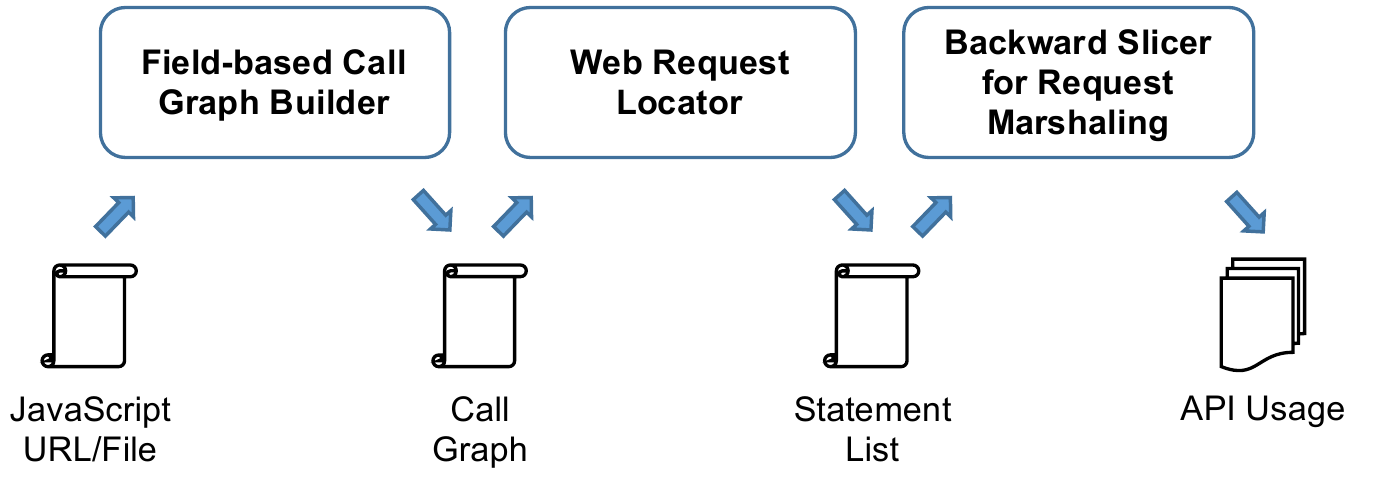}
\caption{Web API Usage Extractor Overview.}
\label{fig:pipeline}
\end{figure}

A fundamental part of our approach is to detect and extract web API usages from JavaScript source code. Figure~\ref{fig:pipeline} shows the web API usage extractor pipeline. The input is a JavaScript file. The output is web API usage including URLs and request payloads in JSON format. The decision to have a single JavaScript file as input, i.e., setting the analysis scope to the file-level as opposed to the program-level, is based on our initial observations that the input strings are often in the same file as the requests. This decision also supports an analysis light-weight enough to be used repeatedly during development. For the rest of this section, we describe the three main components in the pipeline:

\smallskip

\noindent
\textbf{Field-based Call Graph Builder}: The extractor takes a JavaScript file as input and parses the script, excluding files with syntax errors. The analysis then translates the script into the intermediate representation and builds an approximate call graph, called a \emph{field-based call graph}~\cite{Feldthaus_icse2013}. A field-based call graph is a statement-level call graph that uses one abstraction for all instances of each property used in the program, as opposed to one abstraction for each property of each abstract object as in traditional call graphs; this has been shown to scale well for framework-based JavaScript web applications even in the presence of JavaScript's dynamic features~\cite{DBLP:conf/pldi/SchaferSDT13}. In our implementation, we used the field-based call graph construction available in WALA~\cite{WALA}. For optimization, this call graph construction in WALA used to ignore all data flow that does not involve functions. To support our string analysis, we extended the data flow analysis in the call graph construction to also track data flow of strings in the program. We take all functions in the script as entry points for the call graph. Standard approaches take event handlers and top-level blocks as entry points. However, if we were to use the same approach, our analysis with the scope at the file-level could miss entry points if functions are registered as event handlers in a script beyond the analysis scope.

\smallskip

\noindent
\textbf{Web Request Locator}: To identify API invocations, we look for framework-specific patterns in the call graph.  For jQuery, we handle the most common patterns, i.e., function calls to \code{\$.ajax}, \code{\$.get}, and \code{\$.post}. We note instructions that make such calls and use them as the seeds for the inter-procedural data flow analysis, the next component in the pipeline.  When a script does not contain a matched invocation statement, our analysis does not produce any output and the pipeline terminates.

\begin{figure}[t]
\centering
\includegraphics[width=\columnwidth]{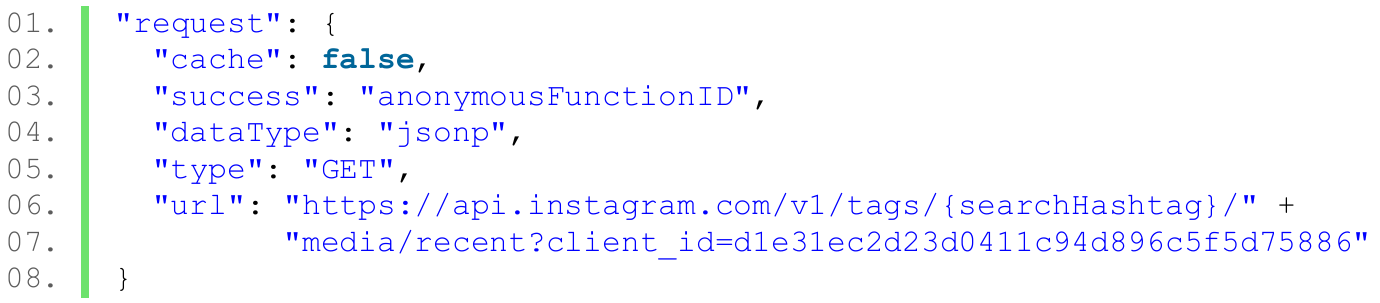}
\caption{Extracted API usage for the example in Figure~\ref{fig:analysis_moti}.}
\label{fig:analysisOutput}
\end{figure}

\smallskip

\noindent
\textbf{Backward Slicer for Request Marshaling}: In this step, we extract the statements that contribute to the input of the web API invocations. Starting from each request function call captured in the previous step, we apply standard inter-procedural backward slicing~\cite{compiler_design_book} to narrow down the subset of statements of the program that affect the statement containing the request. In our implementation, we used the WALA backward slicer~\cite{WALA}. To get the actual strings pertaining to the URL and other parameters in a request, we recover all possible data flows that lead to the request. String values that cannot be determined until runtime are represented by symbolic values (e.g., the value of \code{searchHashtag} in Figure~\ref{fig:analysis_moti} is retrieved from the front-end in line $5$).
% track down variable definitions, and assemble strings and variables. 
% If the value of a variable cannot be determined until run-time, we use a symbolic value. 
For strings with symbolic values, we model common string operators (i.e., concatenations and \code{encodeURI}). For constant strings, we model additional string operators including \code{substring}, \code{replace}, and \code{indexOf}. Currently, research on robust and scalable modeling of string operators with symbolic values is still ongoing \cite{Z3-str, CVC4, Z3str2, S3P}. However, we believe these cases are not significant in the web API usage extraction. This assumption is also supported by the observations in the experiment: we only found two cases where this limitation led the checker to incorrectly flag a string URL as a mismatch to the specification (Section~\ref{sub:endpoints}).

\begin{figure}[t]
    \centering
    \subfloat[{\scriptsize Multiple paths}]{{ \includegraphics[width=\columnwidth]{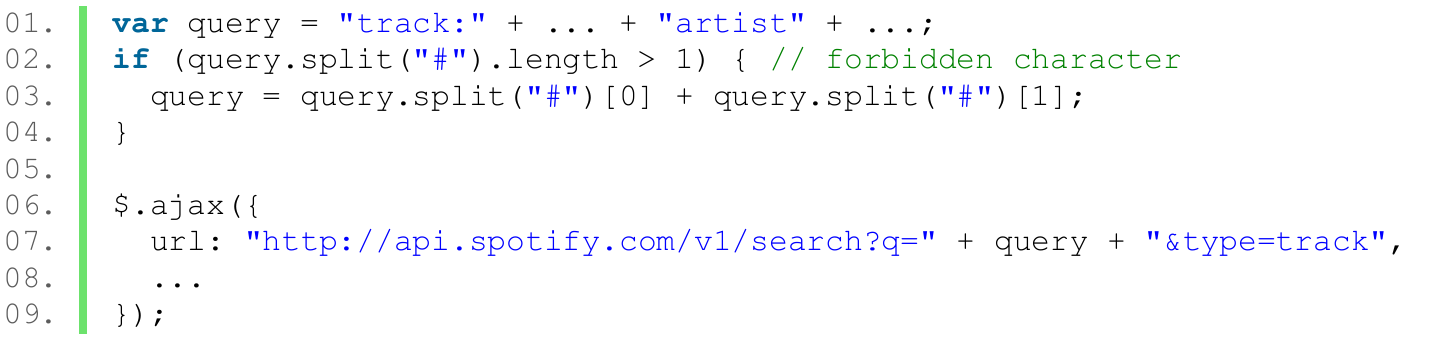}}\label{fig:multiValue1}} \\ \vspace{-0.1in}
    \subfloat[{\scriptsize Multiple callers}]{{ \includegraphics[width=\columnwidth]{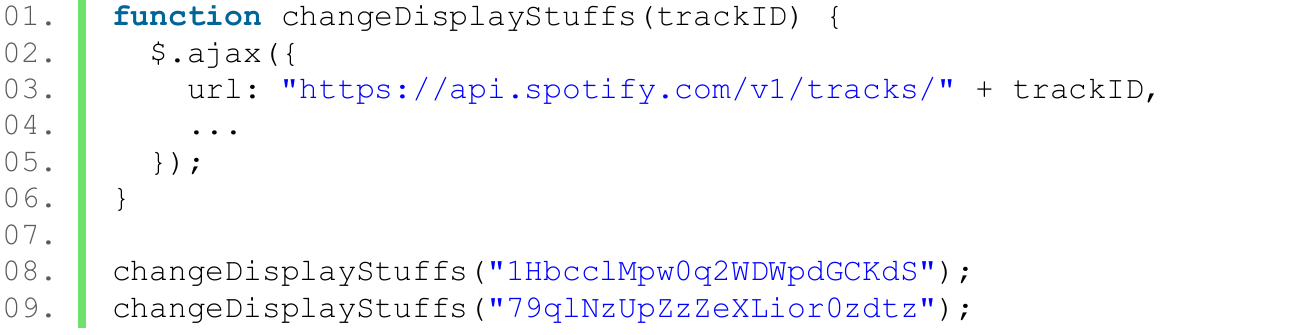} }\label{fig:multiValue2}}
    \caption{Code excerpts exemplifying the extraction of multiple URLs for one request.}
    \label{fig:multiValue}
\end{figure}

We assume all execution paths leading to the request are feasible and we thus perform path-insensitive data flow analysis. It is possible that a variable has multiple definitions from different paths. For example, Figure~\ref{fig:multiValue} shows two common patterns where multiple URLs can be  extracted from a request. In Figure~\ref{fig:multiValue1}, variable \code{query} in line $7$ can have different values depending on the predicate in line $2$.  In Figure~\ref{fig:multiValue2}, function \code{changeDisplayStuffs} can be invoked with different parameter values in line $8$ and $9$. For such cases, we take the union of all possible values. Finally, we output the analysis result in JSON format. The analysis output for the example shown in Figure~\ref{fig:analysis_moti} is presented in Figure~\ref{fig:analysisOutput} as an example. The extracted data contains the retrieved URL and HTTP method as well as all other properties passed to the \code{\$.ajax} function.

% !TEX root = ./paper.tex

\section{Checking Procedure}
\label{sec:checker}
The goal of the checking procedure is to match the information produced by the static analysis against formal web API specifications.
The procedure reveals inconsistencies between the request implementation and the specification.
%, which point to errors either in the one or the other.
In general, the information about each request consists of (1) one URL of the web API to invoke, including the path identifying the endpoint and possibly a query string, (2) the HTTP method, and when required (3) data being sent in the payload body of the request.
Due to the nature of the analysis, though, in practice multiple URL values, HTTP methods, and payload data can be retrieved for a single request because the static analysis considers all possible execution paths to the invocation (as mentioned previously in Section~\ref{sec:approach}).
We designed the checking procedure to compare any possible combinations of these data points against a web API specification.
If any combination matches a specification, no error is reported so as to not annoy users with false positives.

\subsection{Checking endpoints}
\label{sec:checker_endpoints}
The first part of the procedure aims to match a request to an endpoint defined in an API specification.
% Matching API specifications:
The procedure starts by checking whether any of the URLs for a given request begins with any of the base URLs of the known API specifications.
If an API specification contains more than one base URL, for example as it defines schemes \code{HTTP} and \code{HTTPS}, all versions are checked.
Furthermore, it is possible that multiple specifications are found to match to a request, for example because multiple specification versions exist for the same API or because the static analysis reports multiple URLs for the request.
If no specification can be matched, the procedure reports an error, instructing users to check whether the base URL is defined correctly.

% Matching paths:
Next, the procedure attempts to match a request's URLs to paths defined in the API specifications.
To match a path, the procedure takes every URL of a request and compares it against the path definitions in every specification previously matched to that request.
Every path definition of the request's URL strings is retrieved by truncating the base URL defined in the specification and the query string, if it exists.
The remaining path strings are then compared against the path definitions in the specifications by checking whether every path component (separated by \code{"/"}) matches.
This matching considers that both, the path strings from the URLs and the Swagger path definitions, may contain variables denoted by curly brackets.
The procedure treats these path components as wild cards.
Multiple path definitions may be matched to a single request, because multiple specifications can be matched and because a request may contain multiple URLs.
%If no match can be found, the procedure reports an error, pointing to the attempt of invoking a non-existing or deprecated path or to a spelling mistake.

% Matching methods:
Finally, the procedure determines if the HTTP method matches the specification.
If the static analysis does not report an HTTP method, the method is assumed to be \code{GET}.
Any method determined by the static analysis is checked against all methods defined in all matched specifications and all matched paths.
%If no method can be matched, the procedure reports a corresponding error.

\subsection{Checking request data}
\label{sec:checker_req_data}
For requests that can be matched to an endpoint definition in an API specification, the procedure additionally checks the validity of request data (if it exists).
Request data is either the data sent in the payload body of a request (typically of \code{POST}, \code{PUT}, or \code{PATCH} requests) and the data sent within a query string.

\subsubsection{Checking payload data}
\label{sec:checker_req_data_payload}
Data sent in the payload body of an HTTP request can be in any format.
%for example the Extensible Markup Language (XML), YAML Ain't Markup Language, or Comma Separated Values (CSV).
As the static analysis focuses on JavaScript, and because its the prevalent data format in web APIs~\cite{Rodriguez:2016}, we focus on data in the JavaScript Object Notation (JSON).
Swagger specifications allow the expected payload data to be defined, either for certain paths (across all methods) or for specific endpoints (an endpoint-level definition overrules a path-level one).
Payload data definitions can be specified in place, or by referencing definitions in the central \code{definitions} section of the specification.
The procedure considers all these ways to define payloads and, if needed, resolves conflicts of definitions on different levels.
If any of the matched specifications defines a payload schema in any of the matched endpoints,
the procedures determines if the payload data reported in the request information adheres to that schema or not.
A possible violation is that a property marked as \code{required} in the schema is not present in the data.

\subsubsection{Checking query parameters}
\label{sec:checker_req_data_query}
The query data is encoded in key-value pairs.
Within API specifications, query parameters can be defined as either optional or required.
The checking procedure, then, can determine whether all required query parameters are present in a request.
Again, the procedure considers definitions of query parameters from different locations in a specification and to resolve possible conflicts between definitions on different levels.
To check the query parameters, the procedure parses the query strings of all URLs reported for a request.
It then checks whether any of the found parameter sets matches the parameter definitions found in any of the endpoint definitions matched for the request.

% !TEX root = ./paper.tex

\section{Evaluation}
\label{sec:experiment}
To evaluate the web API request checker described in Sections~\ref{sec:approach} and \ref{sec:checker}, we applied it to the problem of identifying and checking whether JavaScript web APIs requests are consistent or inconsistent with a API specifications. The input of the checker is a set of JavaScript code mined from GitHub as well as a set of Swagger specifications (Section~\ref{sub:data}).

For the evaluation, we are interested in two research questions:
\begin{itemize}
  \item \textbf{RQ1}: Given JavaScript code describing a web API request, to what degree can the analysis correctly determine whether the request is consistent with  an endpoint in given Swagger specifications? (See Section~\ref{sub:endpoints})
  \item \textbf{RQ2}: For a request consistent with an endpoint in the Swagger specifications, to what degree can the analysis correctly determine whether the request data (the payload and the query parameters) is consistent with specifications? (See Section~\ref{sec:evaluation_req_data_payload} and Section~\ref{sec:evaluation_req_data_query})
\end{itemize}

For both of these questions, we first determined \emph{quantitatively} how consistent were the information extracted from the static analysis compared to Swagger specifications. To obtain this quantitative information, we count positive matches as well as errors reported from our checking procedure.

For the set of requests that does \emph{not} match endpoints or request data requirements of a specification, we determined if each of the inconsistent instances was legitimate or due to deficiencies in the approach. To ascertain the cause of each of these inconsistencies, we performed a \emph{qualitative} analysis.  
For example, the analysis determined that the endpoint \code{https://api.instagram.com/v1/subscriptions} is inconsistent with the Swagger specification of the Instagram API because the specification does not contain a \code{/subscriptions} path. The qualitative analysis determined that this particular path was deprecated from Instagram's API. With the result of the qualitative analysis, we tabulated the number of instances where the approach correctly identifies endpoints and request data as consistent or inconsistent.

\subsection{Data Collection}
\label{sub:data}
The evaluation requires two types of input data: First, we obtained web API specifications to compare against information produced by the static analysis using the checking procedure. Second, we mined JavaScript source code that (likely) contains requests to the APIs for which we have specifications.

\subsubsection{API specifications from APIs Guru}
\label{sub:data_specs}
%% To obtain a reasonable amount of specifications,
For the Swagger specifications, we made use of a community-maintained collection of specifications from the APIs Guru repository~\cite{APIsGuru}. The repository contains specifications either provided by API providers or third-parties, or generated using dedicated scripts. At the time of performing the experiments, we collected $260$ specifications, which pertain to $230$ APIs (with some APIs having specifications for multiple versions). These specifications act as a source of ground truth to indicate whether requests in the source code invoke the API correctly. We discuss threats to this ground truth in Section~\ref{sec:threats}.

\subsubsection{JavaScript files from GitHub}
\label{sub:data_js}
To increase the generalizability of the results, we aimed to collect a large set of JavaScript files containing web API calls. We obtained such a set by querying GitHub using its search capabilities.\footnote{\url{https://github.com/search}} Each search query targets the domain name of an available Swagger specification from APIs Guru and JavaScript instances that send requests using the jQuery function \code{\$.ajax()} (though we handle additional ways to make requests, i.e., \code{\$.post} and \code{\$get}).
We thus used the search queries of the form \code{extension:js "\$.ajax" "\{domainName\}"}. To automate the data collection, we used Selenium~\cite{Selenium} to invoke the GitHub code search and crawled the search result to obtain links to JavaScript files. The files returned by the search may not necessarily contain requests to the target domain. For example, the domain name may be in a comment and the script still matches the search criteria. From these queries, we obtained $6746$ JavaScript files, from which we extracted $19668$ web API requests. 

For this evaluation, we focused on the $6575$ requests that matched a specification (i.e., a request URL matching the base URL of a specification, including matching the schemes, domain, and the basepath).
We removed the remaining $13093$ requests from our evaluation, 
with $9915$ of the requests were safe to remove:  
These requests neither contained web API calls ($4926$)\footnote{
Of the $4926$ requests, $4177$ requests contained URLs targeting internal endpoints (including relative URLs, localhosts, IP addresses), not web APIs;
$567$ requests only contained empty strings, null values, or numeric values;
and $182$ requests contained URLs matching the domain but not the base path, likely to be targeting static web pages or data.}
nor were the requests the GitHub search intended to target
(i.e., $4989$ requests that did not match the domain of any of the evaluated APIs).
The remaining $3178$ requests corresponded to URLs that could not be resolved by our static analysis, e.g., containing symbolic values in the base URL.  Some of the symbolic values were global variables that could have been resolved if the analysis scope were expanded beyond the file-level,
while other symbolic values were not typically possible to be resolved by static analysis (e.g., values provided to the application at run-time by a user or a configuration file).
The strength of our approach is that our analysis does not raise a false alarm on these $3178$ cases.

%We removed a considerable amount of requests whose URL was reported to be a relative path, which would prohibit us to match the request to an API specification. In addition, the static analysis reported only variables instead of URLs for a large number of requests. This limitation may be caused by scoping the analysis to a single JavaScript file, which prohibits it from resolving global variables - removing this limitation is planned in future work. We further eliminated requests where the reported URLs, though absolute, were considered invalid by the checking procedure. Invalid means, for example, that the URL contains reserved characters in the wrong place (for example, \code{@}) or have an invalid host etc. Finally, we removed around $1300$ requests that contain valid URLs, which do, however, not relate to any of the collected API specifications. The existence of such requests in the data is likely caused by the imprecision of the GitHub code search facilities.

\subsection{Endpoint Results}
\label{sub:endpoints}
For RQ1, the goal was to determine what percentage of the request endpoint URLs extracted from code was correctly flagged as consistent or inconsistent with the Swagger specifications, i.e., the \emph{precision} of the approach. The JavaScript files obtained from GitHub as described in Section~\ref{sub:data} contained $6575$ requests in which the endpoint of a URL matched one of the Swagger specifications, i.e., matches the base URL, including matching the defined schemes (\code{http} or \code{https}), domain, and the basepath.

\begin{figure}[t]
\centering
\includegraphics[width=\columnwidth]{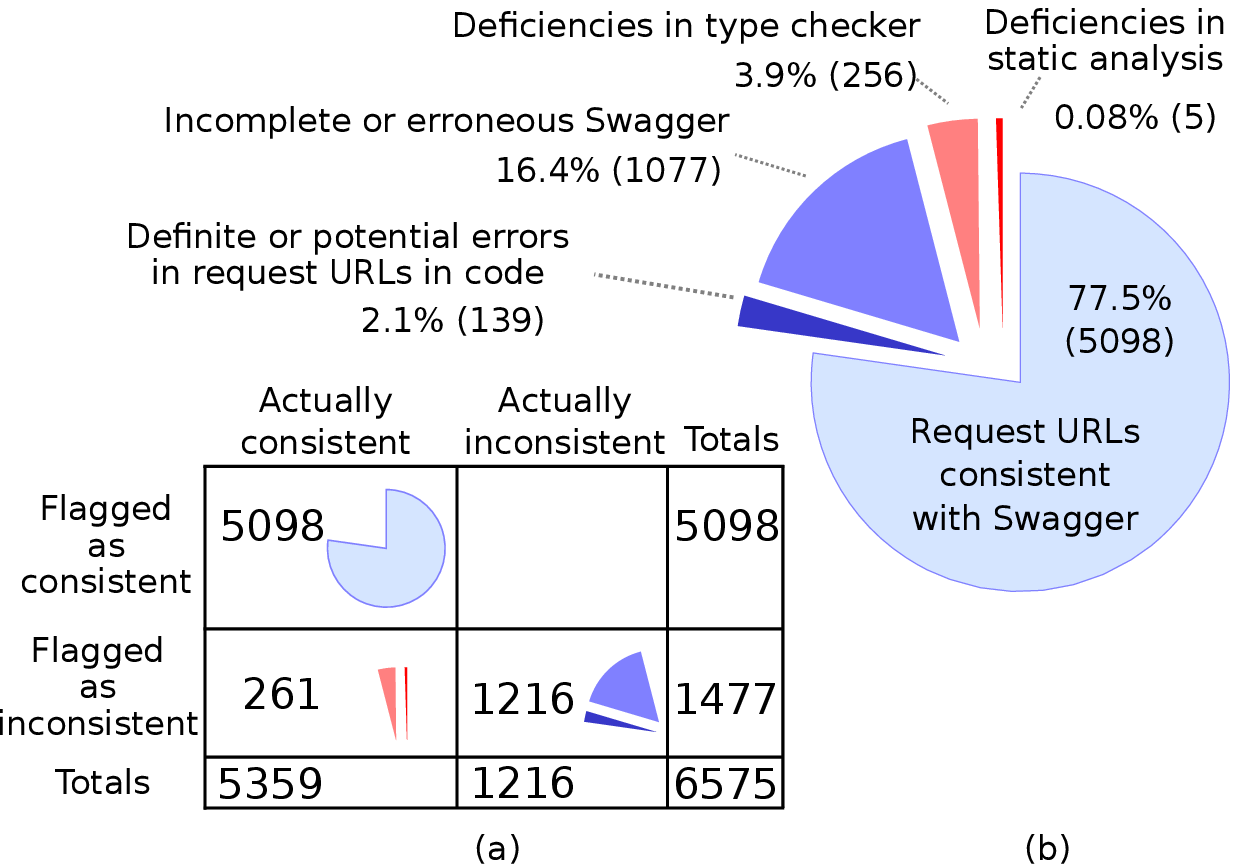}
%pie-prettified}
\caption{Distribution of the $6575$ extracted endpoint invocations}
\label{fig:pie}
\end{figure}

Overall, we found the precision of matching the endpoints of requests to be $96.0\%$, which was tabulated from requests that were flagged consistent and were actually consistent (\includegraphics[scale=0.28]{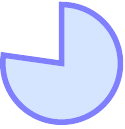} - $5098$ requests in the upper left cell in Figure~\ref{fig:pie}a), and requests that were flagged inconsistent and were actually inconsistent (\includegraphics[scale=0.35]{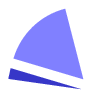} - $1216$ requests in Figure~\ref{fig:pie}a).
% The checking procedure (Section~\ref{sec:checker}) was responsible for automatically determining that $5098$ of the $6575$ requests were consistent with a valid endpoint from a Swagger specification, corresponding to the largest section (\includegraphics[height=0.35cm]{fig/endpoints-true-positives}) of the pie chart in Figure~\ref{fig:pie}b.
In addition to matching the base URL in a Swagger specification, a request is a valid endpoint when it satisfies two conditions: (1) the URL matches an endpoint path, which can contain path variables (e.g., \code{/repos/\{owner\}/\{repo\}}) and (2) the HTTP method matches (e.g., \code{GET}).

For the $1477$ requests that did not match to a valid endpoint (\includegraphics[height=.35cm]{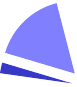} and \includegraphics[height=.35cm]{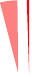}, the ``Flagged as inconsistent'' row in Figure~\ref{fig:pie}a), we qualitatively determined if the static analysis checker correctly identified a legitimate inconsistency or not.
We found that a significant fraction of these requests were true negatives
due to errors in the actual URLs from the JavaScript code ($139$, $2.11$\%)
and incomplete, missing, or erroneous Swagger specifications ($1077$, $16.4$\%),
together pertaining to $1216$ requests (\includegraphics[height=.35cm]{fig/endpoints-true-negatives} in Figure~\ref{fig:pie}a).
The remaining $261$ requests were false positives (\includegraphics[height=.35cm]{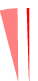} in Figure~\ref{fig:pie}a) due to 
deficiencies in the static analysis ($5$, $0.08$\%) and in the checking procedure ($256$, $3.9$\%).
For the rest of Section~\ref{sub:endpoints} we present the qualitative analysis of the results on these true negatives and false positives.

\cell{Endpoint Results: True Negatives - \includegraphics[height=.4cm]{fig/endpoints-true-negatives}}
Requests that were correctly flagged as inconsistent fall into two categories:
definite or potential errors in the code, and erroneous and incomplete specifications.

For the first category, we found that $139$ requests contained an erroneous URL from the JavaScript code that we correctly flagged as inconsistent, accounting for $2.1$\% of the $6575$ requests (Figure~\ref{fig:pie}b). Of the $139$ requests, $23$ were attributed to deprecated calls and programming errors that were definitely erroneous:

\begin{itemize}
  \item \pie{Deprecated APIs}: $16$ requests were inconsistent because a URL corresponded to a call to an API that were deprecated entirely, or calls to subset of an API that happened to be deprecated.  For example, the request to \code{https://www.googleapis.com/freebase/v1/text/en
  /bob\_dylan}
  was to the Google's Freebase Internet Marketing API, which has been deprecated since June 30, 2015.\footnote{\url{https://developers.google.com/freebase/v1/topic-overview}}

  \item \pie{Documented Programming Errors}:  In two cases, we found evidence (e.g., in the form of a question posted on a forum) that the URLs in the code were erroneously constructed because of errors in interpreting API documentation.\footnote{\url{http://stackoverflow.com/questions/11606101/how-to-get-user-email-from-google-plus-oauth}}

  \item \pie{Typographical errors}: In five cases, the requests contained obvious typographical errors. For example, \code{https://api.spotify.com/v1/seach} did not match the endpoint \code{search} in the Swagger specification of the Spotify API because of the typographical error in ``seach''.  In another case, the checker reported that the extracted URL string
  \code{'https://api.spotify.com/v1/users/'+userID+ /playlists/+playlistID+'/tracks'}
  (a string that looked like a syntax error because of the absence of quotes surrounding \code{/playlists/})
  did not match any endpoints in the Spotify API, even though there was an endpoint \code{http://api.spotify.com/v1/users/\{userId\}/play lists/\{playlistId\}/tracks} in the Spotify's Swagger specification.
  One could argue that the checker flagging this case as a mismatch was a mistake, because at runtime, \code{/playlists/} without the quotes actually evaluates to the string \code{'/playlists/'} as \code{/playlists/} is interpreted as a regular expression.  However, we argue that marking this case as a mismatch is legitimate because it is likely that the author of the code intended to include the quotes \code{'/playlist/'} but this potential error was not caught by the JavaScript interpreter nor testing.
\end{itemize}
 
%% In one case, the URL corresponds to a beta API call documented to expire in a few months.

In addition, we found $116$ requests with potential errors: % ERIK changed from 118 to 116 so that numbers match

\begin{itemize}
  \item \pie{HTTP method}: In $112$ cases, the URL was using the wrong HTTP method, e.g., using GET instead of POST as specified in a Swagger specification.  Of these cases, $74$ were GET requests to \code{https://www.googleapis.com/oauth2/v1/token info} that should be POST according to the Swagger specification and online API documentation, even though the server accepts the GET call.  We categorized these cases as potential errors because even when an API provider may accept such calls, it is still worth-while to warn a programmer that the call is not consistent with the definition of the API. % ERIK changed from 114 to 112 so that numbers match

  \item \pie{Port number}: We observed four requests in which a URL contained port numbers.  These cases may be problematic because port numbers are seldom in any publicly advertised base URLs. For the four cases involving port numbers, one could argue that it was worth issuing a warning to a programmer because port numbers are unlikely to be in a legitimate base URL. With the static checker that could flag that a URL in the requests were consistent with a basepath but not any of an endpoint, a programmer authoring code containing these 155 cases could potentially made aware of these bugs.
\end{itemize}

\smallskip

%Erroneous and Incomplete Specifications
For the second category of the true negatives, we found that $1077$ requests (or $16.4\%$; Figure~\ref{fig:pie}b) corresponded to invocations that were not matched with any endpoint of the corresponding Swagger specifications. From a manual inspection on the documentation on the API being invoked in the requests, we found that the mismatch was due to erroneous and incomplete Swagger specifications:

\begin{itemize}

  \item \pie{Errors in Swagger specifications}: We determined that $867$ cases due to errors in the Swagger specifications. These erroneous requests were because a base path had been refactored to include a version number (e.g., the \code{1.0} path segment in \code{https://mandrillapp.com/api/1.0/messages/ send.json}) but the Swagger specification was not updated.  

  \item \pie{Missing endpoints in Swagger specifications}: For $210$ requests in which a URL found by the static analysis and verified by us as valid calls, a corresponding endpoint definition with the assumed path was missing from the Swagger specifications. These requests corresponded to $18$ endpoints across four APIs: Reactome, Slack, Trello, and Google APIs.

  \item \pie{Missing authorization URLs in Swagger specifications}: Eight URLs relate to endpoints for authentication purposes, e.g., \code{https://slack.com/api/oauth.access} and \code{https://trello.com/1/appKey/generate}. A complete Swagger specification should include such necessary authentication URLs.  However, in these 8 cases those authentication URLs were missing in the specifications.
\end{itemize}

As it turned out, the $1077$ requests flagged by the checker as inconsistent with the Swagger specification were actual errors in the Swagger specification. Upon informing APIs Guru of these errors, we learned that these specifications were already corrected.\footnote{\url{http://www.apiful.io/intro/2016/05/16/challenges-in-maintaining-specs.html}} This scenario demonstrates a potential use case for using the checker on a large repository of API usage for identifying missing or erroneous information on a given set of Swagger specifications.

\cell{Endpoint Results: False Positives - \includegraphics[height=0.5cm]{fig/endpoints-false-positives}}

There were two sources of mistakingly flagging a request to be a mismatch: deficiencies in the static analysis and deficiencies in the checking procedure. For the first source of errors, we found that only five out of the $1496$ cases were due limitations in the static analysis -- which is surprising, given the challenges in analyzing JavaScript and that the analysis scope was within a file. There were two types of deficiencies:

\begin{itemize}
  \item  \pie{Limitation of the analysis scope}: In three cases, because the analysis scope was within a file, the analysis failed to construct a valid URL because the code contains variables or function calls defined outside the file.
  
  \item \pie{Handling string library functions}:  In two cases, the requests used the library function \code{split}. In the static analysis, we explicitly model other more common string operations, i.e., string concatenations and \code{encodeURI}, as we described in Section~\ref{sec:approach}. Currently, handling a more complete set of string library functions with symbolic values is an active research topic~\cite{Z3-str, CVC4, Z3str2, S3P}. We could leverage such research and model more string operators in the future.
\end{itemize}

\smallskip
The few deficiencies caused by the static analysis show that our technique and the chosen analysis scope are feasible for the problem of extracting requests.

\smallskip
Due to deficiencies of the checking procedure, our approach mistakenly determined that $256$ requests were inconsistent with Swagger specifications. There are two main causes of these mistakes:

\begin{itemize}
  \item \pie{Conservative Matching}: We designed to checker be confident in flagging requests as ``consistent with the Swagger'', at the risk of flagging legitimate URLs as potentially inconsistent. In consequence, we found $251$ requests that the procedure mistakenly flagged as inconsistent with any of the Swagger specifications. For example, the code \code{page.config.baseUrl+'tags/'+term+'media/ recent?client\_id='+page.config.clientId} should have matched the endpoint \code{tags/\{tag\}/ media/recent}. However, the checking procedure requires that the variable is constrained to a single path segment, i.e., characters without a '/' in it. % ERIK changed from 249 to 251 so that the numbers add up

  \item \pie{Missing authorization URLs}: In five cases, the requests were for authorization. These URLs are defined in a Swagger file but the checker does not currently check for such URLs.
\end{itemize}

Overall, using the checker for determining whether an invocation of an web API endpoint corresponds to a valid one in a Swagger specification yielded a promising result, with $96.0\%$ of the endpoint invocations correctly flagged as consistent or inconsistent with the Swagger specifications.

% !TEX root = ./paper.tex

\begin{figure}[t]
\centering
\includegraphics[width=\columnwidth]{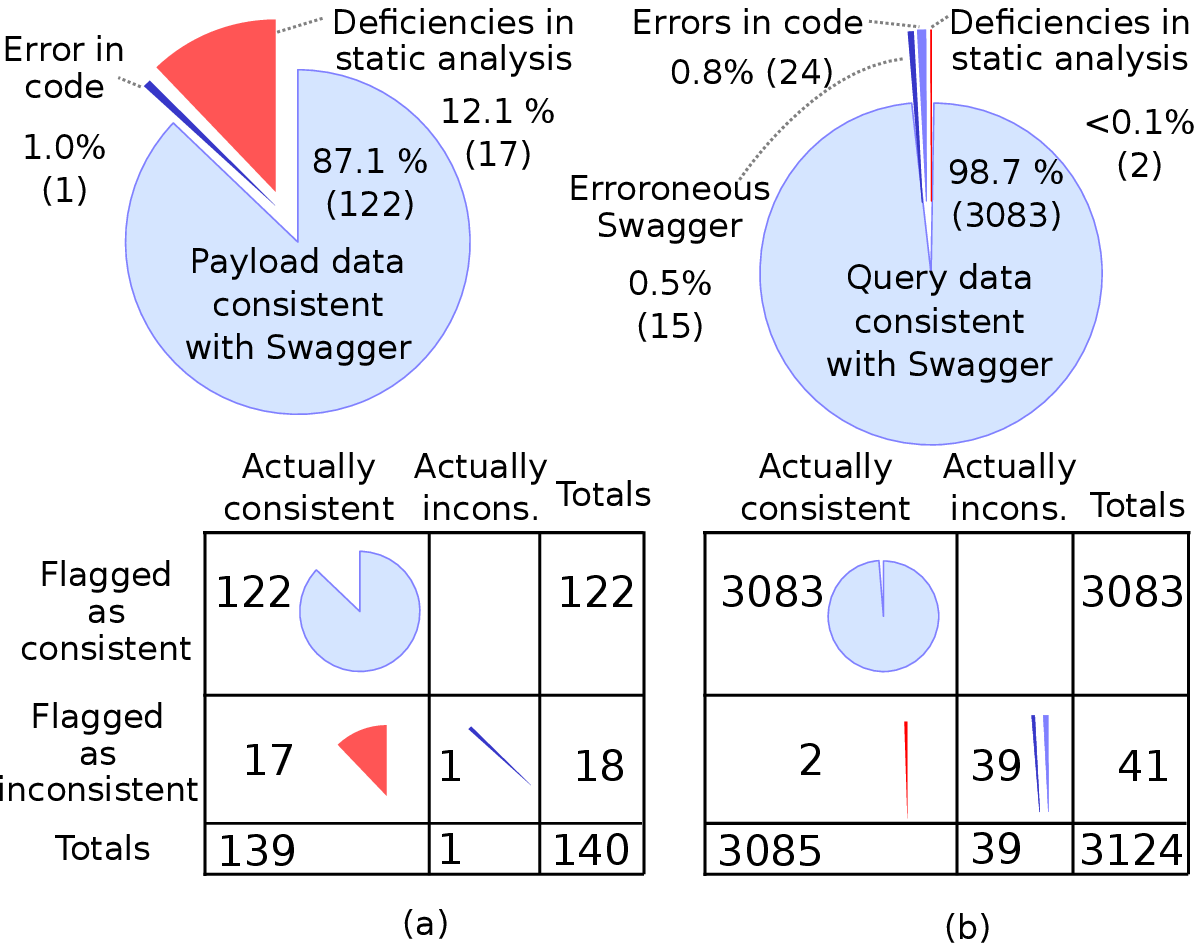}
\caption{RQ2 results for payload (a) and query data (b)}
\label{fig:data-pie}
\end{figure}

\subsection{Payload Data Results}
\label{sec:evaluation_req_data_payload}
For RQ2, we focused on how well the analysis correctly determines whether the request data is consistent with web API specifications. We present the results for the payload data in this Section~\ref{sec:evaluation_req_data_payload} and query data in Section~\ref{sec:evaluation_req_data_query}.

To assess the approach's ability to check for correct payload data, we first determine how many of the $5098$ requests for which we can match an endpoint have a payload schema definition in any of the corresponding API specifications. We only consider requests for which the payload definition is mandatory, i.e., if the request matches multiple specifications or multiple endpoints as described in Section~\ref{sec:checker}, they all need to denote a payload definition. We found that overall $140$ requests have a mandatory payload definition (see Figure~\ref{fig:data-pie}a). Out of these $140$ cases, we found that $122$ requests (\includegraphics[height=0.35cm]{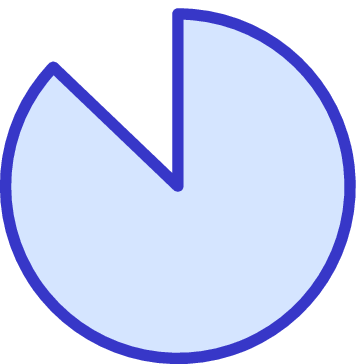}) contained data extracted from the static analysis that adheres to the required payload. Of the $18$ payloads that did not adhere to any of the specifications, we qualitatively ascertained whether the approach correctly determined the mismatches (i.e., true negatives which correspond to  \includegraphics[height=0.35cm]{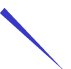} in Figure~\ref{fig:data-pie}a) or not (i.e., false positives which correspond to \includegraphics[height=0.35cm]{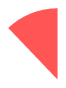}). Tabulating the matched cases (\includegraphics[height=0.35cm]{fig/payload-true-positives.eps}) with the true negatives (\includegraphics[height=0.35cm]{fig/payload-true-negatives.eps}) yields a precision of 87.9\%.

\cell{Payload Data Results: True Negatives ( \includegraphics[height=0.35cm]{fig/payload-true-negatives.eps} )}
The single true negative case is caused by an error in the code: While the analysis reports a required data property to be sent in a query parameter, the specification requires it to be sent in the payload body.

\cell{Payload Data Results: False Positives (\includegraphics[height=0.35cm]{fig/payload-false-positives.eps})}
The $17$ false positive cases are explained by deficiencies in the static analysis. In $13$ cases, the static analysis does actually report data that, upon manual inspection, does match with the data schema in the specification. However, in these cases, the data is present as a JSON-encoded string in the request source code (rather than a JSON object), so our checking procedure fails to correctly match it. In four other cases, the analysis reports a variable to be replaced by a global JSON string, which cannot be resolved statically.

\subsection{Query Parameter Results}
\label{sec:evaluation_req_data_query}
Regarding the query parameters, we first determined how many of the $5098$ requests for which we could match an endpoint have query parameters defined in the API specifications. Like in the case of the payload data, we only considered requests that match an endpoint that has a mandatory query parameter definition, of which we found $3124$. Figure~\ref{fig:data-pie}b presents the breakdown of these requests. We found that $3083$ of the requests (\includegraphics[height=0.35cm]{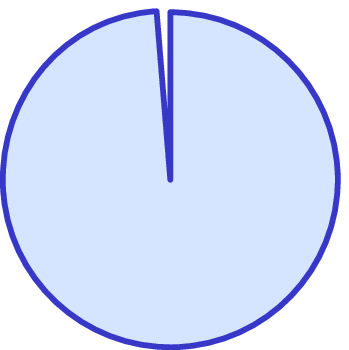}) complied with their corresponding query parameter definitions.  We qualitatively analyzed the $41$ cases in which our approach reported a mismatch, of which $39$ were true negatives ( \includegraphics[height=0.35cm]{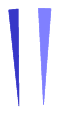} ) and 2 were false positives ( \includegraphics[height=0.35cm]{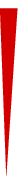}). The precision is 99.9\%, taking into account the matches (\includegraphics[height=0.35cm]{fig/query-true-positives.eps}) and the true negatives ( \includegraphics[height=0.35cm]{fig/query-true-negatives.eps} ).

\cell{Query Parameter Results: True Negatives -  (\includegraphics[height=0.35cm]{fig/query-true-negatives.eps})}
The $39$ cases in which our approach correctly detects an inconsistency between source code and specification fall into two categories:

\begin{itemize}
  \item \pie{Errors in the specification}: % 15 cases
  % GitHub: (15)
  $15$ cases concern \code{GET} requests to the \code{../repos/\{owner\}/{repo}/contributors} path of the GitHub API.
  According to the API specification, a query parameter \code{anon} is required to indicate whether to list anonymous contributors. Invoking this endpoint test-wise reveals that requests also succeed without that parameter in the request, pointing to an error in the specification.

  \item \pie{Errors in the code}: % 24 cases
  We find $24$ cases where required data is sent in the wrong place.
  % Spotify - top tracks: (11 + 5 + 2 = 18)
  In $11$ cases, \code{GET} requests to the \code{../artists/\{id\}/top-tracks} path of the Spotify API miss a \code{country} query parameter, which is required according to the API's specification and online documentation. Interestingly, though, all $11$ requests send a \code{country} data property in their payload body.
  Similar cases can be observed for five \code{GET} requests to the \code{../artists} path and two \code{GET} requests to the \code{../me/following/contains} and \code{../me/tracks/contains} paths, where, again, required query parameters are sent in the payload body instead.
  % Instagram & Slack & Bufferapp: (4 + 1 + 1)
  Similarly, in four \code{POST} requests to the Instagram API, one \code{GET} request to the Slack API, and one \code{GET} request to the Buffer API, query parameters, which are required based on the specification, are sent in the payload body instead.
\end{itemize}

\cell{Query Parameter Results - False Positives ( \includegraphics[height=0.35cm]{fig/query-false-positives.eps} )}
We observe two cases where the static analysis misses to report an inconsistency because the extracted URLs end with a variable. The variable spans across parts of the endpoint path and possibly the query parameters. Thus, in this cases, the analysis fails to report required query parameters.

\smallskip
Overall, out of the $39$ reported query parameter mismatches, $15$ cases are explained by errors in the specification - required parameters are actually not necessary for a successful request. In $24$ cases, query parameters that are required according to the specification are sent in the payload body of the request instead, which is, in most of these cases, in conflict with both, the API specification and online documentation.
Only in two cases do we find mismatches due to our approach's failure to report needed information.

% !TEX root = ./paper.tex

\section{Related Work}
\label{sec:related-work}
Our work on checking JavaScript code with respect to web API specifications is similar in spirit to a range of work on checking and bug finding approaches, such as TypeScript~\cite{Typescript} and JSHint~\cite{JSHint}, respectively. Due to the dynamic nature of JavaScript, and the extreme difficulty of providing precise analysis, such languages and tools tend to be lenient; rather than attempting to be complete, they work by partially enforcing type rules or using a set of patterns that can be tuned to provide some level of feedback without overwhelming their users with a large number of false reports.
Our work shares that approach: our analysis is biased to only report issues that are fairly likely to be real. However, our approach differs from other bug finding tools for JavaScript by being based on inter-procedural static analysis, and relying on traditional techniques such as inter-procedural slicing and string analysis.

% Checking web API specifications:
% Some research addresses the empirics of web API usage.
% A large-scale analysis of monitored, real-life web API requests assesses their characteristics, like the used data format, HTTP method, or whether the payload is well-formed~\cite{Rodriguez:2016}.
% The analysis finds that JSON is the most commonly used data format in web API requests, which supports our decision to focus on it.
% Other works address how web APIs evolve over time~\cite{Espinha:2014}, and, more specifically, how well mobile applications function if confronted with changes in the APIs they use~\cite{Espinha:2015}.
% The revealed problems of applications to deal with web API changes act as a motivation for the here presented work, which could help to detect and fix errors in web API requests.

Halfond et al. introduce static analysis techniques for understanding web API usage in Java applications~\cite{Halfond:2008:AIP:1453101.1453126,Halfond:2009:PII:1572272.1572305}.  One technique focuses on how APIs are used in HTML code that is dynamically generated as part of a Java web application~\cite{Halfond:2008:AIP:1453101.1453126}.  The work focuses on first approximating the HTML and then extracting invocations from it.  This approach works well when the logic creating the request is on the server side; it does not target API calls generated in JavaScript on the client side, which is our focus.  Other work introduces symbolic execution to improve results, once again focusing on Java web application~\cite{Halfond:2009:PII:1572272.1572305}.

In the context of JavaScript, related work has shown that understanding API specifications can make dynamic testing more effective~\cite{Jensen:2013:SID:2491411.2491421}. In contrast to this work, we focus on static analysis rather than on dynamic testing. Our work also relates to checking JavaScript function calls. SAFE\textsubscript{WAPI}~\cite{SAFEWAPI} analyzes JavaScript and checks function calls against the web IDL specifications of those functions. SAFE\textsubscript{Wapp}~\cite{SAFEWAPP} models the web application execution environment (e.g., DOM) and checks function invocations against ECMAScript rules. In comparison, we check web API requests, which is not a language construct in JavaScript. In addition, resolving the targets (endpoints) and parameters requires string analysis.

% Best practices
Multiple works have been proposed to check web APIs for compliance with best-practices~\cite{Palma:2015,Rodriguez:2016}.
These works currently take as input observed web API requests~\cite{Rodriguez:2016} or human-readable documentations~\cite{Palma:2015}, but could be adapted to work on top of API specifications.
In contrast to checking specifications, we here propose to check the adherence of source code with given specifications.

% Inferring web API descriptions:
% Some works address the problem of automatically creating and maintaining web API specifications. 
% One approach generates API specifications from observing HTTP requests to an API using a proxy~\cite{Sohan:2015b}.
% Another work assesses API server logs and tries to determine path parameter and parameter types using machine learning~\cite{Suter:2015}.
% These approaches are complementary to the here presented work: automatically generated specifications can be used by tools implementing the here presented method to advise developers in how to implement API request.

% !TEX root = ./paper.tex

\section{Threats to Validity}
\label{sec:threats}
% No header data
By considering the URL, HTTP method, and request data, our approach covers important parts that determine whether a request contains any errors or not.
However, our approach currently does not examine request headers, which can, for example, contain authentication information that affects the validity of a request.
Despite this limitation, our approach is still a valuable first step towards a wider coverage of errors on web API requests.

% Collection of data from GitHub
A threat to validity of our evaluation is in the way we retrieved source code from GitHub (see Section~\ref{sub:data}).
Our data collection relies on the code search facilities provided by GitHub.\footnote{\url{https://github.com/search}}
GitHub only provides limited insights into the search algorithm, for example, that characters like ``.'', ``/'', or ``\textbackslash'' will be ignored and that only small repositories (less than 384 KB and less than 500,000 files) are indexed.\footnote{\url{https://help.github.com/articles/searching-code/}}

The API specifications we used in our experiments may contain errors, similar to any other type of documentation. However, we consider APIs Guru to be one of the most reliable sources of API specifications. They have a policy in place to validate specifications that are not updated within $48$ hours. In addition, APIs Guru reports the origin of specifications (which typically stem from API providers or the developer community).

% Manual checking of results
Finally, a threat to the validity of our experiments is that we manually analyze the errors reported by our checking method. We performed this analysis to shine a light on the sources of errors and took care to cross-validate among multiple sources as much as possible.

% !TEX root = ./paper.tex

\section{Conclusion}
\label{sec:conclusion}

In this paper, we have leveraged existing research in static analysis scalable to framework-based JavaScript web applications and created an analysis capable of extracting strings pertaining to web APIs requests. We used these extracted request data as input to a checker that determines whether the requests are consistent or inconsistent with formal web API specifications. A qualitative analysis of the results from our checker on $6575$ requests shows that most of reported inconsistencies were due to errors in the client code (calls to deprecated APIs, errors in the URLs, errors in data payload definitions) and incomplete Swagger specifications, as opposed false positives. Quantitatively, we found that the approach can correctly determine whether a request is consistent or inconsistent with web API specification with a high precision of $96.0\%$ for endpoint checking, $87.9\%$ for payload data checking, and $99.9\%$ for query parameter checking.

These results point to the promise in creating tools that are capable of warning programmers of source code containing inconsistent web API requests that can be potentially erroneous. As such, this approach can be integrated with existing tools that support developers in using web APIs~\cite{APIHarmony}. Our proposed checker can also be employed with continuous integration for checking the validity of web API usage in case a web API undergoes changes. Furthermore, our work can lead to tools to help API providers to monitor usages of their APIs in publicly available code or integrate with third-parties change monitoring sites such as API ChangeLog.\footnote{https://www.apichangelog.com/} As for future work on the checker itself, we aim to extend the scope of our static checking method to consider additional aspects of web API requests like header information, HTTP response codes, or the structure of returned data. In addition, based on the positive results with jQuery, we want to extend our implementation to handle other web frameworks.

\newpage
\bibliographystyle{IEEEtran}
\bibliography{IEEEabrv,sigproc}  % sigproc.bib is the name of the Bibliography in this ca

\end{document}